\documentclass[twocolumn,superscriptaddress]{revtex4-2}
\usepackage{amsfonts,amsmath,amssymb}
\usepackage{graphicx}
\usepackage{bm}
\usepackage[ruled]{algorithm2e}
\usepackage{float}
\usepackage{hyperref}
\usepackage[usenames,dvipsnames,svgnames,table]{xcolor}
\usepackage{setspace}
\definecolor{goodblue}{RGB}{0, 91, 187}
\hypersetup{
  colorlinks=true,
  allcolors=goodblue,
  urlcolor=goodblue,
  citecolor=goodblue,
  pdfborder={0 0 0},
  breaklinks=true,
}
\usepackage{quotes}
\usepackage{parskip}

\SetKw{And}{and}
\newcommand{\order}[1]{$\mathcal{O}(#1)$}
\newcommand{\ER}{Erd\"os-R\'enyi }

\begin{document}

\title{Opinion disparity in hypergraphs with community structure}
\author{Nicholas W. Landry}
\email{nicholas.landry@uvm.edu}
\affiliation{Vermont Complex Systems Center, University of Vermont, Burlington, Vermont, USA}
\affiliation{Department of Mathematics and Statistics, University of Vermont, Burlington VT, USA}
\affiliation{Department of Applied Mathematics, University of Colorado at Boulder, Boulder, Colorado, USA}

\author{Juan G. Restrepo}
\email{juanga@colorado.edu}
\affiliation{Department of Applied Mathematics, University of Colorado at Boulder, Boulder, Colorado, USA}

\date{\today}

\begin{abstract}

The division of a social group into subgroups with opposing opinions, which we refer to as {\it opinion disparity}, is a prevalent phenomenon in society. This phenomenon has been modeled by including mechanisms such as opinion homophily, bounded confidence interactions, and social reinforcement mechanisms. In this paper we study a complementary mechanism for the formation of opinion disparity based on higher-order interactions, i.e., simultaneous interactions between multiple agents. We present an extension of the planted partition model for uniform hypergraphs as a simple model of community structure and consider the hypergraph SIS model on a hypergraph with two communities where the binary ideology can spread via links (pairwise interactions) and triangles (three-way interactions). We approximate this contagion process with a mean-field model and find that for strong enough community structure, the two communities can hold very different average opinions. We determine the regimes of structural and infectious parameters for which this opinion disparity can exist and find that the existence of these disparities is much more sensitive to the triangle community structure than to the link community structure. We show that the existence and type of opinion disparities are extremely sensitive to differences in the sizes of the two communities.
\end{abstract}

\keywords{hypergraph, community structure, polarization, stochastic block model, planted partition model, epidemic threshold}

\maketitle

\section{\label{sec:introduction} Introduction}

Modeling the spreading and dynamics of opinions on social networks is a problem of major interest in modern network science. Many classical models of opinion dynamics such as the voter model \cite{clifford_model_1973}, the majority rule model \cite{galam_minority_2002,chen_majority_2005}, and the Sznajd model \cite{sznajd-weron_opinion_2000} inevitably result in consensus \cite{hassani_classical_2022}, i.e., the same opinion shared by all agents. In reality, however, consensus is rarely reached; instead, strong differences in opinion between different groups, i.e., {\it polarization}, are commonly observed. This is evident in politics \cite{andris_rise_2015}, social media \cite{cinelli_echo_2021}, and ideology \cite{noauthor_political_2014}. Many opinion models giving rise to polarization are based on opinion homophily, i.e., the tendency of individuals to associate with similar minded individuals \cite{baron_social_1996,macy_polarization_2003, gilbert_blogs_2009, dandekar_biased_2013,gargiulo_role_2017}, bounded confidence interactions \cite{deffuant_mixing_2000,hegselmann_opinion_2005, del_vicario_modeling_2017}, unfollowing on social media platforms \cite{sasahara_social_2021}, and social reinforcement and feedback mechanisms \cite{banisch_opinion_2019, leonard_nonlinear_2021}. Other models for the formation of polarization include the effects of cooperation and partisanship \cite{kawakatsu_interindividual_2021,yang_why_2020}, echo chambers on social media \cite{cinelli_echo_2021}, radicalization dynamics \cite{baumann_modeling_2020}, cognitive biases \cite{wang_public_2020}, moderates seen as outsiders \cite{yang_falling_2021}, and media influence \cite{brooks_model_2020}.

A closely related phenomenon to polarization occurs when social groups have different average opinions, which we refer to as {\it opinion disparity} to avoid confusion with the extensive literature and existing definitions of polarization. In this paper we explore a  mechanism for the formation of opinion disparity driven by the presence of simultaneous interactions between multiple agents, known as {\it higher-order interactions}. Ref.~\cite{iacopini_simplicial_2019} noted that higher-order interactions in the Susceptible-Infected-Susceptible (SIS) model can lead to bistability, hysteresis, and explosive transitions in random simplicial complexes, and Ref.~\cite{landry_effect_2020} extended this result to more complex hypergraphs. Since higher-order interactions lead to bistability in all-to-all and random hypergraphs, one expects that hypergraphs with strong community structure retain --- to some extent --- the ability to sustain different opinions in different communities. Confirming this, Ref.~\cite{ferraz_de_arruda_multistability_2023} found that hypergraph community structure combined with higher-order interactions can result in multistability, intermittency, and complex dynamics for a generalized SIS model on hypergraphs. Ref.~\cite{noonan_dynamics_2021} studied the majority rule model on hypergraphs with community structure and found that for sufficiently disconnected communities the average states of each community are well-separated. Extending these previous results, here we explore the interplay of community structure and higher-order interactions systematically by first developing a generative model for hypergraphs with tunable community structure, and then studying a social contagion model on hypergraphs with community structure to determine which combinations of hypergraph and contagion parameters support communities with different opinions. By analyzing a mean-field model of this social contagion process, we find that opinion disparity is possible for strong enough community structure in the higher-order interactions, and can appear suddenly as the strength of the community structure is increased. Furthermore, we find that the presence of opinion disparity can be extremely sensitive to the relative size of the communities.

The structure of our paper is as follows: we define the contagion model and generative hypergraph models that we use in Section \ref{sec:models}, we use these definitions in Section~\ref{sec:mf_analysis} to develop and analyze mean-field models of contagion spread on hypergraphs with community structure, and we discuss our results in Section~\ref{sec:discussion}.

\section{\label{sec:models} Models}

In this section, we present both a generative hypergraph model that incorporates community structure as well as the opinion model that we use in the rest of the paper.

\subsection{\label{sec:notation} Notation}

A hypergraph, $H=(V, E)$, is a mathematical object that encodes relationships between arbitrary numbers of entities, where $V$ is the set of $N=|V|$ vertices and $E$ is the set of hyperedges, where a hyperedge $e\in E$ is a subset of the vertices. An $m$-uniform hypergraph is a hypergraph where $|e| = m$ for every hyperedge $e$ in the set of hyperedges $E$. The $m$th-order degree of a node $i$, $k_i^{(m)}$, is the number of $m$-hyperedges of which node $i$ is a member. In the following we will use $k = k^{(2)}$ to signify the $2$nd-order degree and $q = k^{(3)}$ to signify the $3$rd-order degree. We will indicate their mean values by $\langle k \rangle$ and $\langle q\rangle$, respectively. We will also refer to hyperedges of sizes $2$ as {\it links} and to hyperedges of size $3$ as {\it triangles}. When an m-uniform hypergraph is specified and there is no possibility of confusion, we will denote $k=k^{(m)}$ to simplify notation.

\begin{figure}
    \centering
    \includegraphics[width=8.6cm]{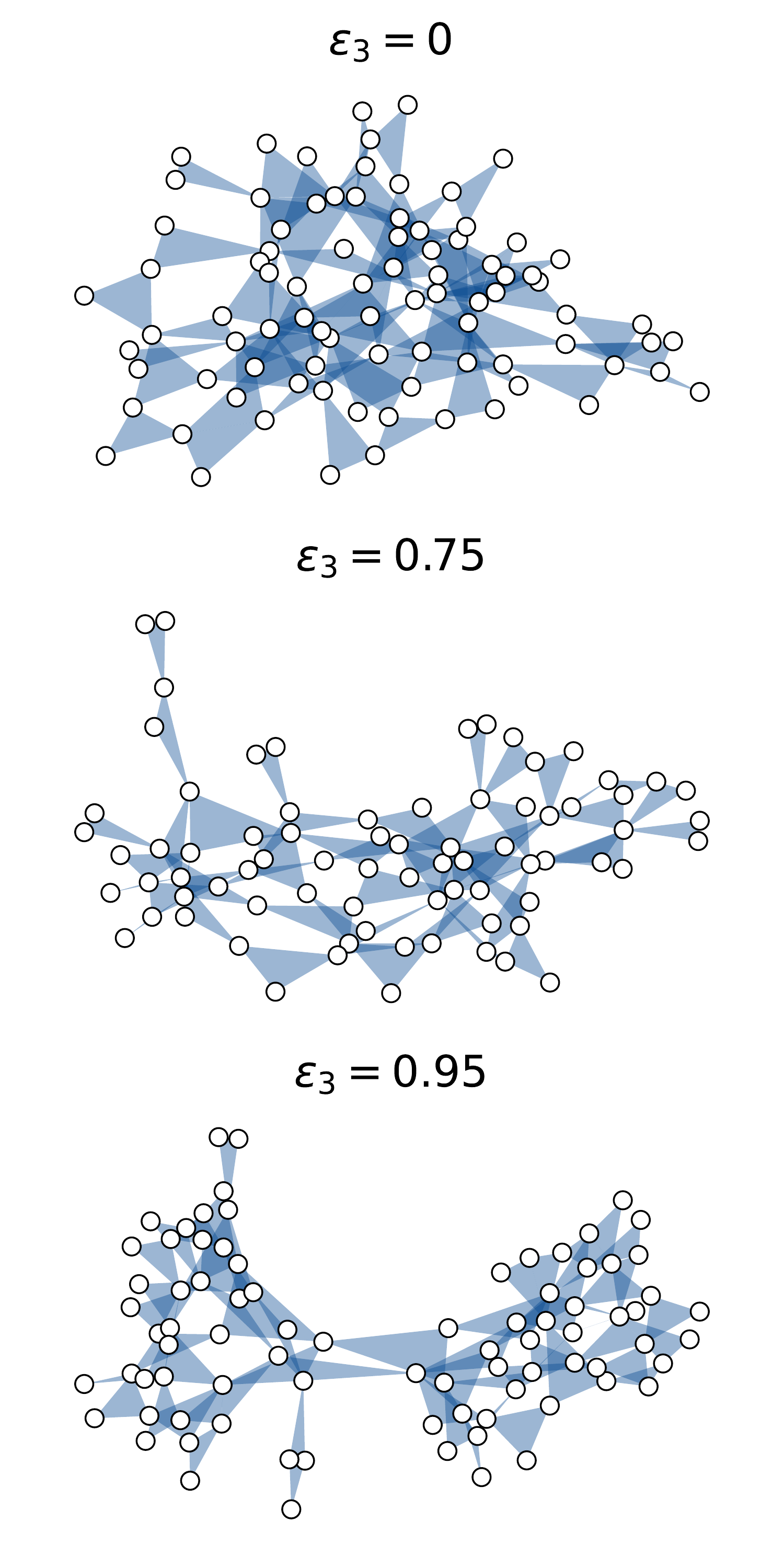}
    \caption{An illustration of a 3-uniform hypergraph sampled from the HPPM model with varying strengths of community structure using XGI \cite{landry_xgi_2023}.}
    \label{fig:illustration}
\end{figure}

\subsection{\label{sec:hsbm} The stochastic block model for uniform hypergraphs}

The stochastic block model (SBM) is a simple random network model incorporating community structure \cite{holland_stochastic_1983}. Given a network with $G$ communities, each node $i$ is assigned a community label $g_i \in \{1,2,\dots, G\}$. The probability that two nodes $i$ and $j$ are connected with an edge is assumed to depend only on their community labels $g_i$ and $g_j$. Extending this notion to $m$-uniform hypergraphs gives rise to the $m$-uniform hypergraph stochastic block model (m-HSBM) \cite{kim_stochastic_2018,cole_exact_2020}. Given an $m$-uniform hypergraph, suppose that each node $i$ has a community label $g_i$ and that the number of communities is given by $G$. Then, given nodes $i_1, \dots, i_m$ with community labels $g_{i_1}, \dots, g_{i_m}$, we define the probability that a hyperedge connects these nodes as $P_{g_{i_1}, \dots, g_{i_m}}$. Because every vertex order in a hyperedge is equivalent, $P$ is symmetric, i.e., $P_{i_1,\dots,i_m}=P_{\sigma_{i_1},\dots,\sigma_{i_m}}$, where $\sigma$ is any permutation of the indices. 

In our numerical experiments, we will generate hypergraphs from a version of this model. Sampling from this model by iterating through every potential edge and accepting it with a specified probability is inefficient for sparse hypergraphs. In Appendix~\ref{sec:appendix_hsbm_algorithm}, we present a more efficient algorithm for sampling sparse hypergraphs from this model. 

\subsection{\label{sec:ppm} Planted partition model for uniform hypergraphs}

Extending the {\it planted partition model} for pairwise networks \cite{jerrum_metropolis_1998} and the hypergraph stochastic block model \cite{hickok_bounded-confidence_2022,ghoshdastidar_consistency_2017,kim_stochastic_2018,cole_exact_2020}, here we introduce the {\it m-uniform Hypergraph Planted Partition Model} (HPPM). Using a single parameter, this generative model interpolates from an m-uniform hypergraph with completely random connections to an m-uniform hypergraph with two disconnected communities, while maintaining a constant expected degree for each node (see Fig.~\ref{fig:illustration}). 

Consider a set $V$ of $N$ vertices divided into $2$ communities of sizes $\rho N$ and $(1-\rho)N$, with $0 < \rho < 1$. For a given set $e$ of $m$ nodes, we let the probability that a hyperedge connects the edges be
\begin{align}
P(e \in E) = \left\{
\begin{array}{cc}
p_{\text{in}}, & \text{all nodes in $e$ belong to }\label{eq:p_def}\\
&\text{ the same community },\\
&\\
p_{\text{out}}, & \text{ otherwise }.\\
\end{array}
\right.
\end{align}
 Note that we are following the "All-or-Nothing" (AoN) definition in Ref.~\cite{chodrow_generative_2021}, which considers a hyperedge to connect two different communities if it contains at least one member of each community. One can relax this assumption by accounting for the fraction of nodes in each community \cite{chodrow_generative_2021,skardal_multistability_2023}, but we do not consider this case here.

In order to create hypergraphs with a tunable amount of community structure, we change the relative size of $p_{\text{in}}$ and $p_{\text{out}}$ while ensuring that the mean degree $\langle k^{(m)}\rangle$, which henceforth will be indicated with $\langle k \rangle$, is constant. When there is no community structure, both probabilities are equal to the hypergraph \ER probability $p_{\text{ER}}$,
\begin{align}
p_{\text{out}} = p_{\text{in}} = p_{\text{ER}} = \langle k \rangle N/ \left[m \binom{N}{m}\right] \approx \langle k \rangle \frac{(m-1)!}{N^{m}},\label{eq:p_er_balanced}
\end{align}
where here and in the following we assume $N \gg m$ and approximate $\binom{N}{m}$ by $N^m/m!$. To determine $p_{\text{in}}$ and $p_{\text{out}}$ when there is community structure, we note that the total expected number of hyperedges $M$ is given by the sum of intra- and inter-community hyperedges
\begin{align}
M = &\binom{\rho N}{m} p_{\text{in}} + \binom{(1-\rho) N}{m} p_{\text{in}} +\nonumber\\
&\left[ \binom{N}{m} -\binom{\rho N}{m} -\binom{(1-\rho) N}{m} \right]p_{\text{out}}.
\end{align}
Using $\langle k \rangle = m M/N$, approximating $\binom{N}{m}$ by $N^{m}/m!$, and using $\langle k \rangle (m-1)!/N^{m-1} = p_{\text{ER}}$, we find
\begin{align}
p_{\text{ER}} &= q_{\rho,m} p_{\text{in}}
+ \left[ 1 - q_{\rho,m} \right]p_{\text{out}}, \label{eq:p_er_imbalanced}
\end{align}
where 
\begin{align}
q_{\rho,m} = \rho^m +(1-\rho)^m.
\end{align}
Now we introduce the {\it imbalance parameter} $\epsilon_{\rho,m}$ \cite{jerrum_metropolis_1998} that interpolates between a hypergraph without community structure ($p_{\text{out}} = p_{\text{ER}}$) for $\epsilon_{\rho,m} = 0 $ and a hypergraph with completely disconnected communities ($p_{\text{out}} = 0$) for $\epsilon_{\rho,m} = 1$. Accordingly, we parameterize $p_{\text{in}}$ and $p_{\text{out}}$ as
\begin{align}
p_{\text{out}} &= p_{\text{ER}}(1-\epsilon_{\rho,m}),\label{pout}\\
p_{\text{in}} &= p_{\text{ER}}(1+ r_{\rho,m} \epsilon_{\rho,m}).\label{pin}
\end{align}
Inserting this into Eq.~\eqref{eq:p_er_imbalanced}, we find
\begin{align}
r_{\rho,m} = \frac{1 - q_{\rho,m}}{ q_{\rho,m}},
\end{align}
which is the ratio between inter- and intra-community edges. This sheds light on why the imbalance parameter in \eqref{pin} is multiplied by $r_{\rho,m}$: since there are more potential inter-community hyperedges than intra-community hyperedges, this factor is needed so that the mean degree remains constant as the relative size of $p_{\text{in}}$ and $p_{\text{out}}$ is changed.

For the special case of equal-sized communities, $\rho = 1/2$, the probabilities simplify to 
\begin{align}
p_{\text{in}} &= p_{\text{ER}} + (2^{m-1}-1)\epsilon_m p_{\text{ER}},\label{pinequal}\\
p_{\text{out}} &= p_{\text{ER}}-\epsilon_m p_{\text{ER}}.\label{poutequal}
\end{align}
Note that in this case, the factor $r_{1/2,m} = 2^{m-1} -1$ can be understood by the fact that, given a node in one community, there are $2^{m-1}-1$ more potential inter-community links than intra-community links. In the following, we define $\epsilon_m \equiv \epsilon_{1/2,m}$ for simplicity. Also in this case, in terms of the expected intra- and inter-community mean degrees
\begin{align}
\langle k_{\text{in}}\rangle &= \binom{N/2 - 1}{m-1} p_{\text{in}},\\
\langle k_{\text{out}} \rangle &=\left[ \binom{N}{m-1} - \binom{N/2 - 1}{m-1}\right] p_{\text{out}},
\end{align}
the imbalance parameter is given by
\begin{align}
\epsilon_m =\frac{\langle k_{in}\rangle - \langle k_{out}\rangle}{(2^{m-1}-1)\langle k\rangle}.
\end{align}

For a given value $\rho$ and $\langle k \rangle$, changing the probabilities of intra- and inter-community links according to Eqs.~\eqref{pout}-\eqref{pin} produces hypergraphs with varying amounts of community structure. Figure~\ref{fig:illustration} shows three hypergraphs obtained by the HPPM with $m = 3$, $\rho = 1/2$, $\langle q \rangle = 2$, and values of $\epsilon_3$ equal to $0$ (top), $0.75$ (middle), and $0.95$ (bottom). 

Before moving on, we note that the HPPM differs from the generative model used in Ref.~\cite{noonan_dynamics_2021} in that the total expected number of hyperedges in the hypergraph is held constant as the amount of community structure is varied, allowing us to isolate the effect of community structure on the dynamics.

\subsection{\label{sec:opinion_model} Opinion model and opinion disparity}

Following Refs.~\cite{landry_effect_2020,iacopini_simplicial_2019}, we study the hypergraph \textit{collective contagion model}. This model allows nodes to hold binary opinions: susceptible ($0$) and infected ($1$). An infected node spontaneously transitions to the susceptible state at a rate $\gamma > 0$ independently of the states of neighboring nodes. A susceptible node may transition to the infected state at a rate $\beta_m$ if that node is a member of a group of size $m$ and all of the other members in this group are infected. It is assumed that the node may be infected independently by each such group it belongs to. We note that the two opinion states are asymmetric in this model: in the absence of influence or infection by other groups, an individual will almost surely heal given enough time; however, a healthy individual will not become infected without the influence of other individuals. Therefore, the opinion corresponding to the infected state may only be sustained if the opinion is continually shared with neighbors. This is akin to a peer pressure effect where a behavior will likely die out without the influence of groups to sustain it.

In Refs.~\cite{landry_effect_2020,iacopini_simplicial_2019}, it was found that this model admits bistable regions where a state with no infection and a state with a macroscopic fraction of infected nodes are simultaneously stable. Therefore, viewing hypergraphs with strong community structure as perturbations of disconnected hypergraphs, we expect that, depending on the parameters, we will find stable solutions where the fractions of infected nodes in each community are very different. 

\begin{figure}[t]
    \centering
    \includegraphics[width=8cm]{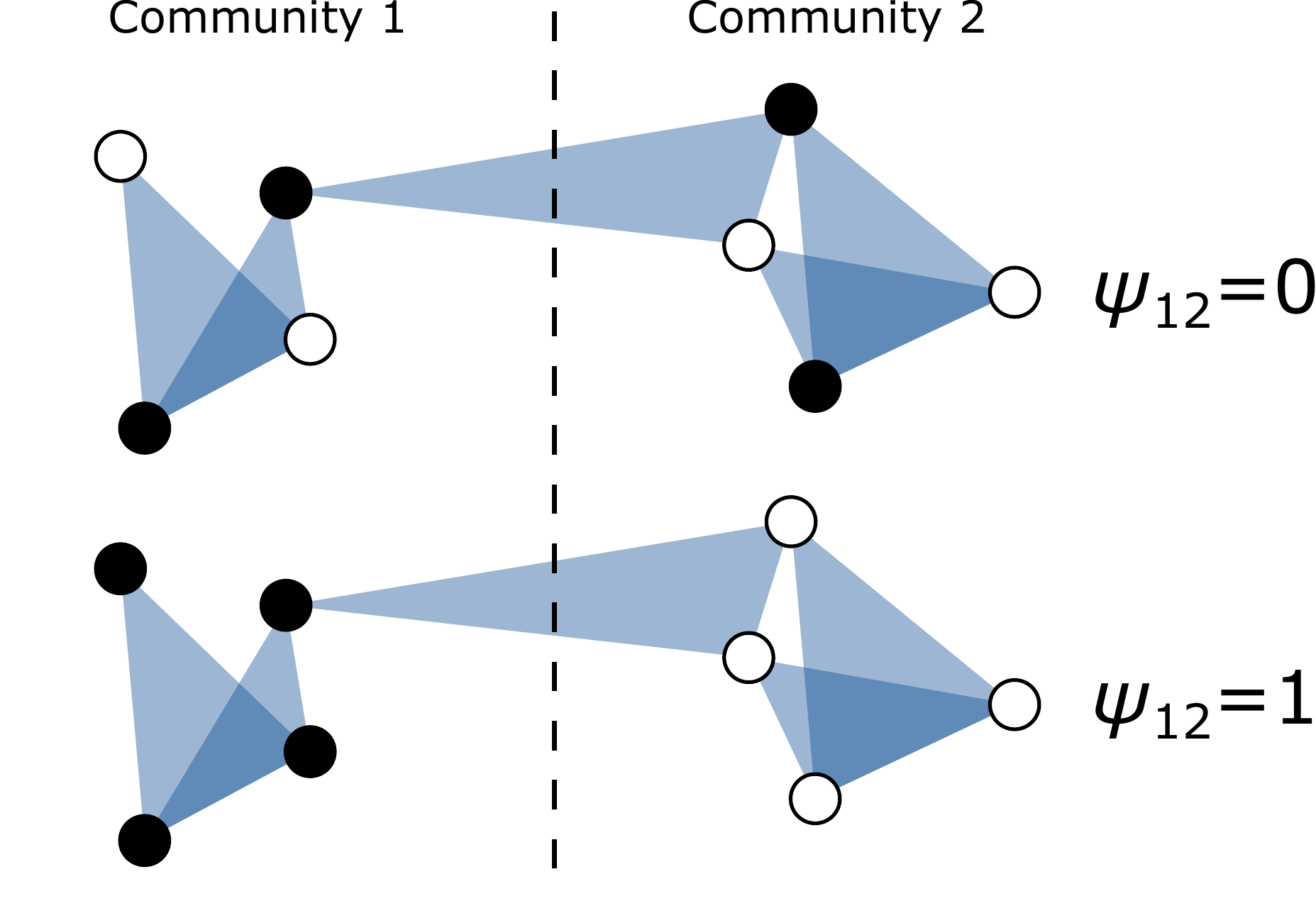}
    \caption{Example of two communities with no opinion disparity (top) and maximum opinion disparity (bottom). Infected and susceptible nodes are shown as black and white, respectively.}
    \label{fig:psidef}
\end{figure}

The observation above motivates the definition of opinion disparity we will use in this paper. Given a hypergraph with community structure, we define the opinion disparity, $\psi_{ij}$, between communities $i$ and $j$ as the maximum difference in average opinions between the two communities that can be sustained at equilibrium. For example, a hypergraph where all the nodes in one community have opinion $1$ and all the nodes in the other community have opinion $0$ in the steady state has the maximum possible opinion disparity, $\psi_{ij} = 1$. On the other hand, when the fraction of nodes with opinion $1$ is the same in each community, the opinion disparity is zero. Note that this definition of opinion disparity assumes the existence of predefined communities, and that the community membership of each node is known. The opinion disparity differs from polarization in that polarization is often measured as a property, such as bimodality, of the opinion distribution \cite{adams_mathematical_2022,dimaggio_have_1996}. In this formalism, bimodal opinion distributions can have zero opinion disparity, depending on the community membership of the nodes (see Fig.~\ref{fig:psidef}). Therefore, our definition is more appropriate to quantify the extent to which observed community structure in a hypergraph is correlated with nodal opinions.
Caution is advised when community structure cannot be reasonably expected because the value of the opinion disparity is dependent on the nodal community labels. In this paper, we do not focus on the inference of these labels, but instead refer the reader to articles discussing higher-order community detection, such as Refs.~\cite{chodrow_generative_2021,cole_exact_2020,chodrow_nonbacktracking_2023,kaminski_clustering_2019}.

\section{\label{sec:mf_analysis} Mean-field analysis}

In this section, we derive and analyze mean-field equations describing the dynamics of the opinion model introduced in Sec.~\ref{sec:opinion_model} for hypergraphs with community structure. For simplicity, we only consider hypergraphs with hyperedges of sizes 2 (links) and 3 (triangles). First, we consider the case of equal sized, i.e., balanced, communities.

\subsection{\label{sec:mf_ppm} Balanced communities}

Consider a hypergraph of size $N$ with hyperedges of sizes 2 (links) and 3 (triangles) assigned following the HPPM in Section \ref{sec:ppm} with parameters $\rho = 1/2$, $\epsilon_{2}\equiv \epsilon_{1/2,2}$ and $\epsilon_{3}\equiv \epsilon_{1/2,3}$, respectively. We denote the fraction of infected individuals in community $i$ as $x_i$ and, assuming $N$ is large, extend the mean-field model of Ref.~\cite{iacopini_simplicial_2019} to include community structure. 

We derive a rate equation for the fraction of infected nodes in the first community, $x_1$, of the form 
\begin{align}
\frac{d x_1}{dt}& = -\gamma x_1 + (1-x_1)[\beta_ 2 N_2 + \beta_3 N_3],
\end{align}
where $-\gamma x_1$ is the rate of spontaneous healing, $(1-x_1)$ is the fraction of nodes that are susceptible, and $N_2$, $N_3$ are the expected number of infected links and triangles, respectively, that a randomly chosen node in community $1$ belongs to. The corresponding equation for $x_2$ can be obtained by permuting $x_1$ and $x_2$.

The number of infected contacts via pairwise interactions, $N_2$, can be obtained by adding the expected number of infected neighbors in communities $1$ and $2$. Note that, according to the planted partition model for $m=2$, the probability that two nodes are connected to each other is $p_{\text{in}} = (1 + \epsilon_2)\langle k\rangle/N$ if they belong to the same community (in this case, community $1$) and $p_{\text{out}} = (1 - \epsilon_2)\langle k\rangle/N$ if they belong to different communities [cf.~Eqs.~\eqref{pinequal} and \eqref{poutequal}]. Multiplying these by the expected number of infected nodes in community $i$, $x_i (N/2)$ and adding, we find
\begin{equation*}N_2 = \frac{1}{2}\langle k\rangle[(1 + \epsilon_2) x_1 + (1 - \epsilon_2)x_2].
\end{equation*}
To find the number of infected triangles to which a node in community $1$ belongs we proceed similarly. According to the planted partition model for $m=3$, the probability that three nodes form a 3-hyperedge is $p_{\text{in}} = 2(1 + 3\epsilon)\langle q\rangle/N^2$ if they are in the same community and $p_{\text{out}} = 2 (1 - \epsilon) \langle q\rangle/N^2$ if they are in different communities. Adding together the cases where the two other nodes are both in community $1$, in communities $1$ and $2$, in communities $2$ and $1$, or both in community $2$, and taking into account that the number of unique triangles in each community including a given node is $\binom{N/2}{2}\approx N^2/8$, we find
\begin{equation*}
N_3 = \frac{1}{4}\langle q\rangle [(1 + 3\epsilon)x_1^2 + 2(1 - \epsilon)x_1 x_2 + (1 - \epsilon)x_2^2].
\end{equation*}
Putting all the terms together, we find
\begin{align}
\frac{d x_1}{dt}&=-\gamma x_1 + \frac{\beta_2}{2} \langle k \rangle (1-x_1)\left[(x_1+x_2) + \epsilon_2(x_1-x_2)\right]\nonumber\\
 + & \frac{\beta_3}{4} \langle q\rangle (1-x_1)\left[(x_1 +x_2)^2 + \epsilon_3(3x_1^2-2x_1x_2 - x_2^2)\right].\nonumber
\end{align}
Setting $\gamma = 1$ (after, if necessary, rescaling time), defining $\tilde \beta_2 = \beta_2 \langle k \rangle $, $\tilde \beta_3 = \beta_3 \langle q \rangle $, and including the analogous equation for $dx_2/dt$, we obtain the system of equations
\begin{align}
\frac{d x_1}{dt}& = -x_1 + \frac{\widetilde{\beta}_2}{2} (1-x_1)\left[(x_1+x_2) + \epsilon_2(x_1-x_2)\right]\label{eq:mf1_rescaled}\\
& + \frac{\widetilde{\beta}_3}{4}(1-x_1)\left[(x_1 +x_2)^2 + \epsilon_3(3x_1^2-2x_1x_2 - x_2^2)\right],\nonumber\\
\frac{d x_2}{dt}& = -x_2 + \frac{\widetilde{\beta}_2}{2} (1-x_2)\left[(x_1+x_2) + \epsilon_2(x_2-x_1)\right]\label{eq:mf2_rescaled}\\
& + \frac{\widetilde{\beta}_3}{4}(1-x_2)\left[(x_1 +x_2)^2 + \epsilon_3(3 x_2^2 - 2 x_1 x_2 - x_1^2)\right].\nonumber
\end{align}
This system of equations characterizes the dynamics of the average opinions in communities $1$ and $2$, $x_1$ and $x_2$. We refer to stable fixed points with $x_1 \neq x_2$ as {\it asymmetric} fixed points, and to fixed points with $x_1 = x_2$ as {\it symmetric } fixed points. According to our definition of opinion disparity in Sec.~\ref{sec:opinion_model}, the existence of stable asymmetric fixed points corresponds to nonzero opinion disparity, given by
\begin{align}
\psi_{12}=\max\{|x_1 - x_2| \text{ such that $(x_1, x_2)$}\label{eq:opinion_disparity}\\
\text{ is a stable fixed point of \eqref{eq:mf1_rescaled}-\eqref{eq:mf2_rescaled}}\nonumber\}.
\end{align}
Since we will consider only two communities, for simplicity we will denote the opinion disparity by $\psi \equiv \psi_{12}$. 
Before discussing the regimes where opinion disparity exists, it is convenient to study the dynamics without opinion disparity, i.e., along the invariant line $x_1 = x_2 = x$. Along that line, the dynamics are given by
\begin{align}
\frac{d x}{dt} = -x + \widetilde{\beta}_2 (1-x)x+\widetilde{\beta}_3(1-x)x^2,\label{eq:x}
\end{align}
which describes the system without community structure studied in Ref.~\cite{iacopini_simplicial_2019}. Eq.~\eqref{eq:x} leads to the symmetric fixed points (corresponding to Eq.~(4) in \cite{iacopini_simplicial_2019})
\begin{align}
x &= 0,\label{zero}\\
x &= \frac{\widetilde{\beta}_3-\widetilde{\beta}_2\pm\sqrt{\left(\widetilde{\beta}_2 + \widetilde{\beta}_3\right)^2 - 4 \widetilde{\beta}_3}}{2 \widetilde{\beta}_3}.\label{nonzero}
\end{align}

\begin{figure}[t]
    \centering
    \includegraphics[width=5.8cm]{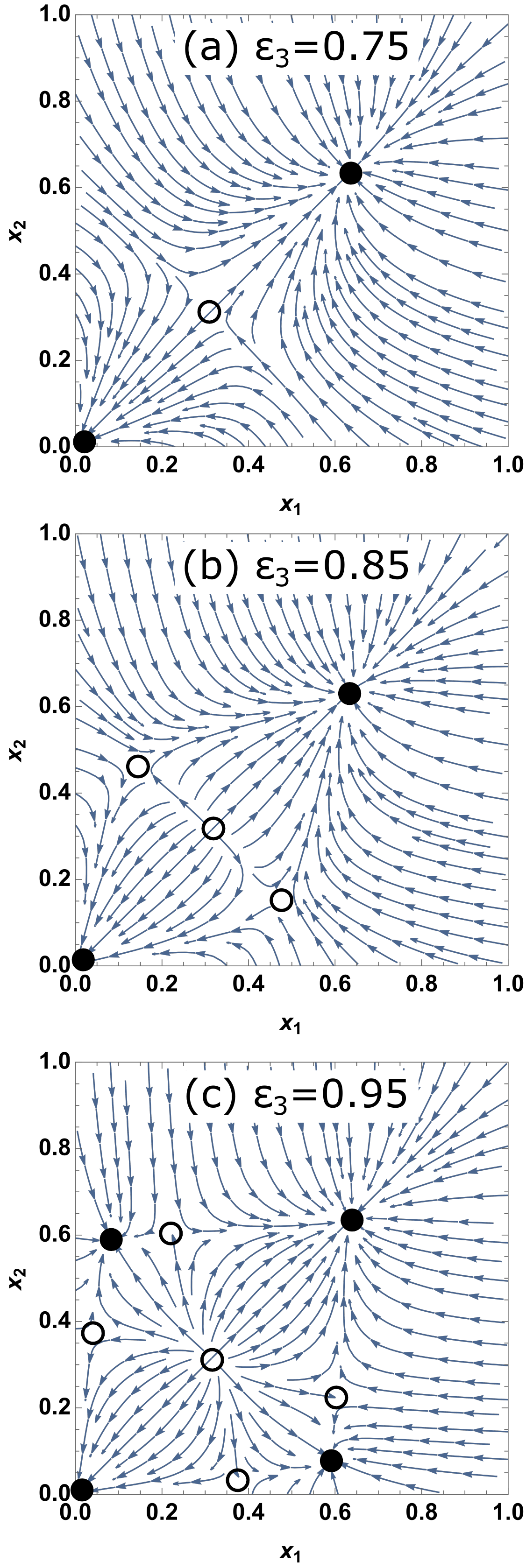}
    \caption{A phase plot of Eqs.~\eqref{eq:mf1_rescaled}~-~\eqref{eq:mf2_rescaled} with $\epsilon_2 = 0.5$, $\widetilde{\beta}_2 = 0.2$, and $\widetilde{\beta}_3 = 4$ for three different values of $\widetilde{\epsilon}_3$. These plots illustrate the cases where there are no asymmetric fixed points ($\epsilon_3=0.75$), only unstable asymmetric fixed points ($\epsilon_3=0.85$), and stable asymmetric fixed points ($\epsilon_3=0.95$). The filled circles correspond to stable fixed points and the open circles correspond to unstable fixed points.}
    \label{fig:phase_diagram}
\end{figure}

\begin{figure*}[t]
    \centering
    \includegraphics[width=14cm]{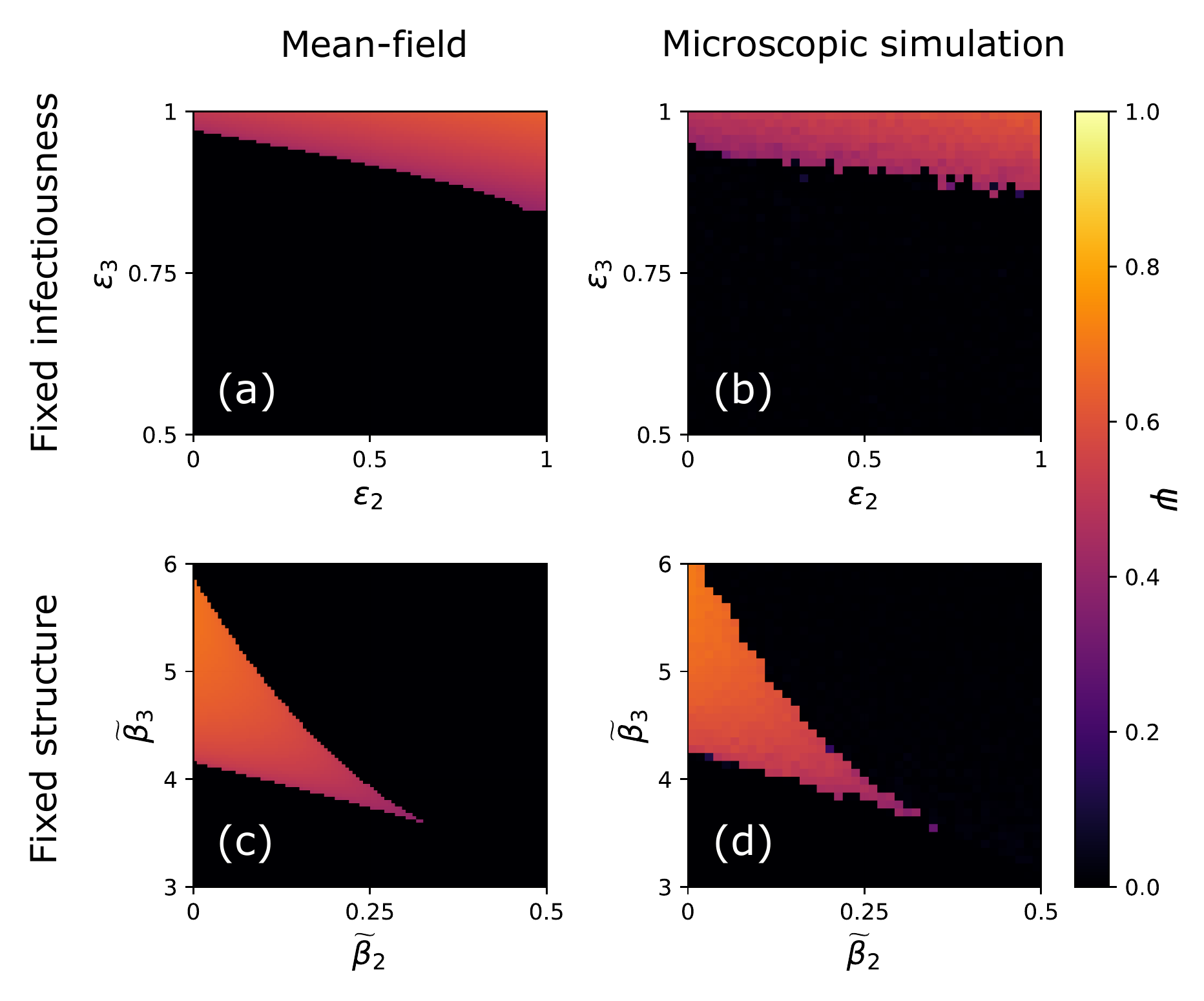}
    \caption{The opinion disparity with respect to community structure (first row) and infectiousness (second row). Note the different ranges of $\epsilon_2$ and $\epsilon_3$ in panels (a) and (b) and $\widetilde{\beta}_2$ and $\widetilde{\beta}_3$ in panels (c) and (d). The opinion disparity is computed in two ways: from the stable fixed points of Eqs.~\eqref{eq:mf1_rescaled}-\eqref{eq:mf2_rescaled} (first column) and from stochastic simulations of the contagion dynamics (second column).}
    \label{fig:psi_panel}
\end{figure*}

Considering now the full dynamics, linearization of Eqs.~\eqref{eq:mf1_rescaled}-\eqref{eq:mf2_rescaled} about the fixed point $(0,0)$ shows that this fixed point becomes unstable for values of $\widetilde{\beta}_2$ larger than the epidemic threshold $\widetilde{\beta}_2^c = 1$. For $\widetilde{\beta}_3>1$, the two nonzero symmetric fixed points (\ref{nonzero}) appear at a saddle-node bifurcation at $\widetilde{\beta}_2 = 2 \sqrt{\widetilde{\beta}_3}-\widetilde{\beta}_3$, and the smaller of these fixed points merges with $x=0$ in a subcritical bifurcation as $\widetilde{\beta}_2$ approaches $\widetilde{\beta}_2^c = 1$. Interestingly, these critical values do not depend on $\epsilon_2$ and $\epsilon_3$. Therefore, this analysis indicates that when the communities have the same size, community structure doesn't modify the \ER model epidemic threshold or onset of bistability, at least regarding the symmetric fixed points. Earlier studies corroborate this result for pairwise networks \cite{volz_effects_2011,stegehuis_network_2021}. However, we note that more accurate approximations that include pair correlations (e.g., Ref.~\cite{burgio_network_2021}) have shown that the epidemic threshold in hypergraph SIS models can depend on the strength of higher-order interactions. Therefore we anticipate that there might be corrections to our results due to correlations not included in the mean-field analysis. 

Having studied the symmetric fixed points, now we turn to the more interesting asymmetric fixed points that support opinion disparity. As an illustrative example, consider the case $\epsilon_2 = 0.5$, $\epsilon_3 = 0.75$, $\widetilde{\beta}_2 = 0.2$, and $\widetilde{\beta}_3 = 4$ shown in Fig.~\ref{fig:phase_diagram}(a). In this case, the only fixed points are the three symmetric fixed points (\ref{zero})-(\ref{nonzero}), and therefore there is no opinion disparity. 
Changing $\epsilon_3$ to $0.85$ results in the same symmetric fixed points and two additional asymmetric unstable fixed points [Fig.~\ref{fig:phase_diagram}(b)] though a pitchfork bifurcation. Increasing $\epsilon_3$ to $0.95$, two additional stable (solid black circles) and two unstable (empty circles) asymmetric fixed points are created through saddle-node bifurcations, resulting in positive opinion disparity [Fig.~\ref{fig:phase_diagram}(c)]. Thus, we see that there is 
a sudden onset of opinion disparity as $\epsilon_3$ is increased above a certain threshold. This behavior is not present for the hypergraph contagion model on random null models without community structure or higher-order interactions.

In the remainder of this section, we will study how the opinion disparity depends on the community structure and infectiousness parameters $(\epsilon_2,\epsilon_3,\widetilde{\beta}_2,\widetilde{\beta}_3)$ for the case of equal-sized communities.
\begin{figure*}[t]
    \centering
    \includegraphics[width=\linewidth]{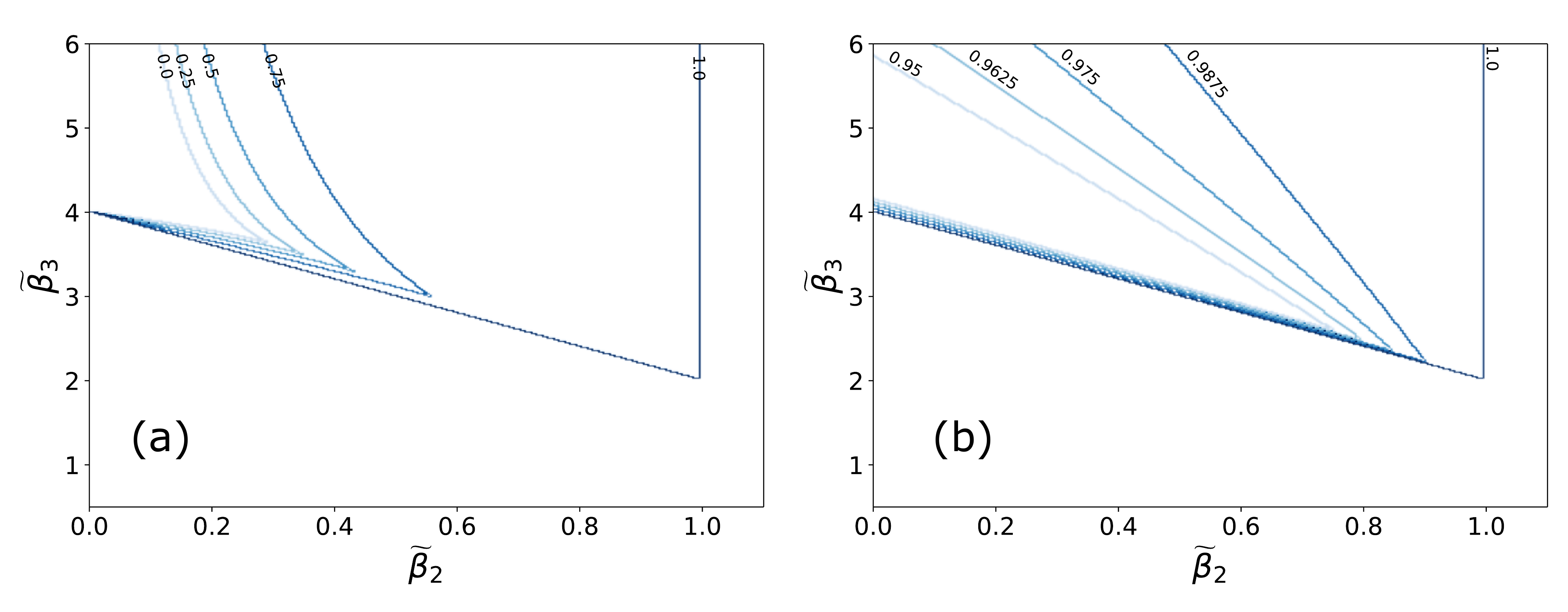}
    \caption{A phase diagram illustrating how changes with (a) $\epsilon_2$ and (b) $\epsilon_3$ affect the regions over which opinion disparity can occur. In this diagram, the upper left region enclosed by each line corresponds to the $\widetilde{\beta}_2,\widetilde{\beta}_3$ region in which opinion disparity can occur for given values of $\epsilon_2$ and $\epsilon_3$, and the remaining area is where opinion disparity does not occur. The boundary lines in panel (a) correspond to $\epsilon_2 = 0, 0.25,0.5,0.75,1.0$ with $\epsilon_3 = 1$ and the boundary lines in panel (b) correspond to $\epsilon_3=0.95,0.9675,0.975,0.9875,1.0$ with $\epsilon_2 = 1$.}
    \label{fig:polarization_boundaries}
\end{figure*}
Given $\epsilon_2$, $\epsilon_3$, $\widetilde{\beta}_2$, and $\widetilde{\beta}_3$, we compute the opinion disparity numerically by setting the derivatives in Eqs.~\eqref{eq:mf1_rescaled}-\eqref{eq:mf2_rescaled} equal to zero, using interval root finding methods to find all fixed points of these 2D coupled equations in $[0,1]^2$, and lastly, selecting the stable fixed point with the largest value of $\vert x_1 - x_2\vert$. The stability of each fixed point is determined from the eigenvalues of the Jacobian associated to Eqs.~\eqref{eq:mf1_rescaled}-\eqref{eq:mf2_rescaled}, shown in Appendix~\ref{sec:appendix_stability}.

In Fig.~\ref{fig:psi_panel}(a), we plot the opinion disparity $\psi$ calculated as described above as a function of $\epsilon_2$ and $\epsilon_3$ with $\widetilde{\beta}_2 = 0.2$ and $\widetilde{\beta}_3 = 4$. We see that there is a pronounced region for large $\epsilon_3$ where opinion disparity occurs. The existence of opinion disparity is much more sensitive to community structure of the triangles ($\epsilon_3$) than that of the links ($\epsilon_2$). To validate our results, in Fig.~\ref{fig:psi_panel}(b) we plot the opinion disparity found from numerical simulation of the infection model on a hypergraph with links and triangles assigned following the HPPM with $m = 2, \langle k\rangle = 20$ and $m = 3, \langle q\rangle = 20$, respectively, and $N = 10^4$. Further details on the numerical simulations may be found in Appendix~\ref{sec:appendix_simulations}. The agreement between mean-field theory and the stochastic simulations is reasonable. One possible explanation for discrepancies between the two is that although opinion disparity may occur in the mean-field equations according to the definition in Eq.~\eqref{eq:opinion_disparity}, in practice, the fixed point might be too weakly stable to allow sustained opinion disparity to occur under the presence of finite-size effects [cf. Fig.\ref{fig:imbalanced_phase_plots}(a)].

Now we look at the effect of the infection rates for a fixed community structure. We expect opinion disparity to be related to the bistable regime of a single community, which occurs when $\widetilde{\beta}_3 > 1$, $2 \sqrt{\widetilde{\beta}_3}-\widetilde{\beta}_3 <\widetilde{\beta}_2 < 1$. If the communities were completely disconnected, this bistable region would correspond to stable opinion disparity. In practice, connections between the communities result in a much smaller opinion disparity region as shown in Figs.~\ref{fig:psi_panel}(c) and (d). In Fig.~\ref{fig:psi_panel}(c), we plot $\psi$ for $\widetilde{\beta}_2$ on the interval $[0, 0.5]$ and $\widetilde{\beta}_3$ on the interval $[3, 6]$ for $\epsilon_2 = 0.5$ and $\epsilon_3 = 0.95$ found numerically from the mean-field equations, and in Fig.~\ref{fig:psi_panel}(d), we show the corresponding microscopic simulations using the same HPPM described above. In Fig.~\ref{fig:polarization_boundaries}, we explore this further by showing how the opinion disparity region shrinks as $\epsilon_2$ (a) and $\epsilon_3$ (b) are decreased from $1$ to $0$ and $0.95$, respectively, when starting from two completely disconnected communities.

In summary, the analysis of the balanced communities shows that opinion disparity can occur for large values of the triangle community structure parameter $\epsilon_3$, and that relatively few connections between the communities can significantly reduce the size of the region in $(\widetilde{\beta}_2,\widetilde{\beta}_3)$ space where opinion disparity occurs. In the next section we will study the case of imbalanced (i.e., different sized) communities. As we will see, even a small difference in size can cause dramatic changes in the opinion disparity.

\subsection{\label{sec:imbalanced_communities} Imbalanced communities}

\begin{figure}[b]
    \centering
    \includegraphics[width=8.6cm]{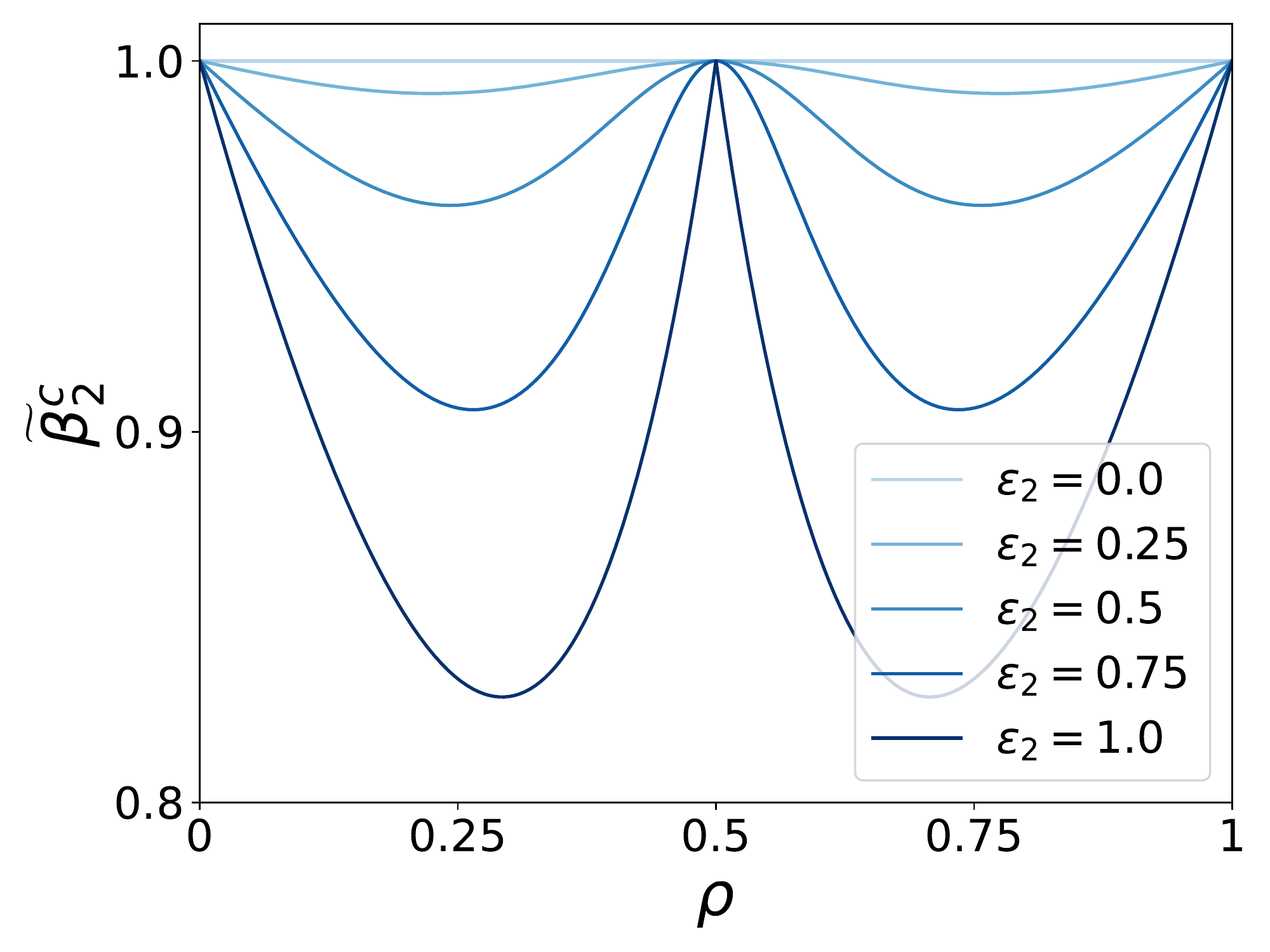}
    \caption{A plot of the epidemic threshold $\widetilde{\beta}_2^c$ with respect to $\rho$ predicted with Eq.~\eqref{eq:beta2c_vs_rho_case2}. The line corresponding to $\epsilon_2=0$ is the epidemic threshold for the \ER case.}
    \label{fig:beta2_vs_rho_case2}
\end{figure}

We now explore the effect of unequal community sizes. Considering the HPPM model of Sec.~\ref{sec:ppm} and performing the same calculations as in Section~\ref{sec:mf_ppm} for an arbitrary $\rho$ we obtain
\begin{widetext}
\begin{align}
\frac{d x_1}{dt} =& -x_1 + \widetilde{\beta}_2 (1-x_1) [\rho(1 + r_{2}\epsilon_{2}) x_1 + (1-\rho)(1 - \epsilon_{2})x_2]\label{eq:mf_imbalanced1}\\
& + \widetilde{\beta}_3 (1-x_1)[\rho^2(1 + r_{3}\epsilon_{3})x_1^2 + 2\rho(1-\rho)(1 - \epsilon_{3})x_1x_2 + (1-\rho)^2(1 - \epsilon_{3})x_2^2],\nonumber\\
\frac{d x_2}{dt} =& -x_2 + \widetilde{\beta}_2 (1-x_2)[(1-\rho)(1 + r_{2}\epsilon_{2}) x_2 + \rho(1 - \epsilon_{2})x_1]\label{eq:mf_imbalanced2}\\
& + \widetilde{\beta}_3 (1-x_2)[(1-\rho)^2(1 + r_{3}\epsilon_{3})x_2^2 + 2(1 - \rho)\rho(1 - \epsilon_{3})x_2x_1 + \rho^2(1 - \epsilon_{3})x_1^2]\nonumber,
\end{align}
\end{widetext}
where we let $\epsilon_{2} = \epsilon_{\rho,2}$, $\epsilon_{3} = \epsilon_{\rho,3}$, $r_2 = r_{\rho,2}$, and $r_3 = r_{\rho,3}$. As before, we can linearize this system and compute the eigenvalues of the Jacobian at $x_1 = x_2 = 0$ to determine stability. The eigenvalues are given by
\begin{equation*}
\lambda = -1 +\frac{1}{2}\widetilde{\beta}_2( 1 + r_2 \epsilon_2\pm\sqrt{\Delta}),
\end{equation*}
where
\begin{align*}
\Delta &= 1+2 \epsilon_2[4 (\rho -1) \rho +(1-2 \rho )^2 r_{2}]\\
&+\epsilon_2^2 [(1-2 \rho )^2 r_{2}^2 - 4 (\rho - 1)\rho].
\end{align*}
Setting the maximal eigenvalue equal to zero and substituting the definition of $r_{2}$ determines the epidemic threshold
\begin{align}
\widetilde{\beta}_2^c &= \frac{2}{1 + \sqrt{1 + \frac{4b}{a}\epsilon_2 - \frac{4c}{a^2}\epsilon_2^2} + (1/a - 1)\epsilon_2},\label{eq:beta2c_vs_rho_case2}
\end{align}
where $a = 1 - 2\rho+2\rho^2$, $b = \rho(\rho - 1)$ and $c = 3\rho^4 - 6\rho^3 + 4\rho^2 - \rho$ are functions of $\rho$. When $\rho = 1/2$ we obtain $a=1/2$, $b = -1/4$, and $c = -1/16$, and we recover the epidemic threshold $\widetilde{\beta}^c_2 = 1$. For all other values of $\rho$, however, the epidemic threshold depends on $\epsilon_2$. To illustrate this, in Fig.~\ref{fig:beta2_vs_rho_case2} we plot $\widetilde{\beta}_2^c$ as a function of $\rho$ for $\epsilon_2 = 0,0.25,0.5, 0.75$, and $1$. In agreement with our previous calculations, the epidemic threshold is independent of $\epsilon_2$ for balanced communities. This plot also shows that different relative sizes are optimal in the spread of a network SIS contagion, in the sense that they result in a lower epidemic threshold. Finally, we note that the epidemic threshold is independent of $\epsilon_3$ for all values of $\rho$.

\begin{figure}
    \centering
    \includegraphics[width=8.6cm]{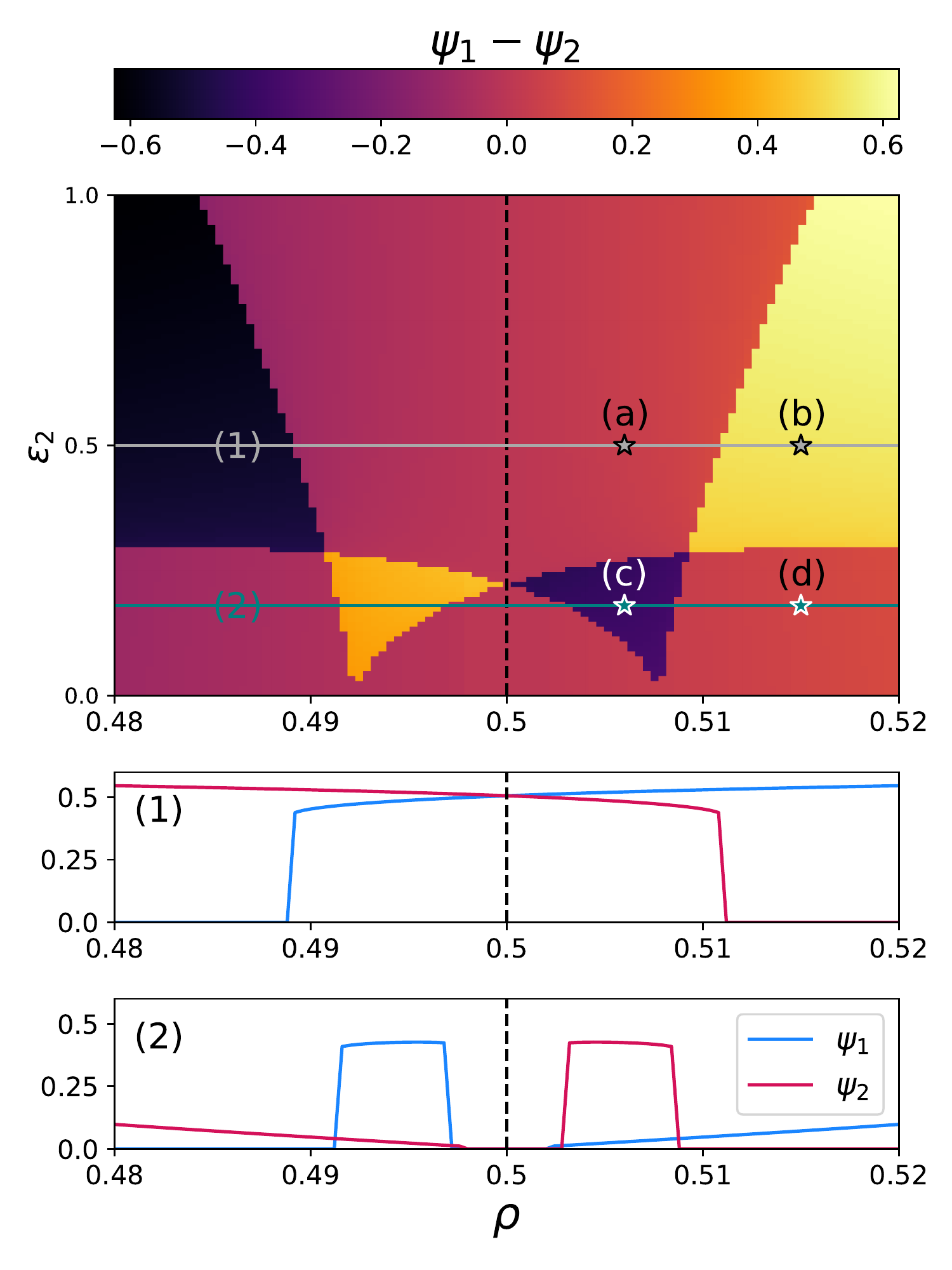}
    \caption{In this plot, $\langle k\rangle = \langle q\rangle = 20$, $\epsilon_3 = 0.95$, $\gamma=1$, $\beta_2 = 0.2/\langle k\rangle$, and $\beta_3 = 4/\langle q\rangle$. The top panel illustrates the difference $\psi_1 - \psi_2$ as a function of $\rho$ and $\epsilon$. The two solid horizontal lines (1) and (2) correspond to the panels below which plot cross-sections of $\psi_1 - \psi_2$, evaluated at fixed $\epsilon_2 = 0.18$ and $\epsilon_2 = 0.5$ values. The points on each cross-section are at $\rho=0.506$ and $\rho=0.515$. $\psi_1 - \psi_2$ is an odd function about the dashed line defined by $\rho = 1/2$. For large enough $\epsilon_2$ as seen in cross-section (1), when community 1 is larger than community 2 (i.e., $\rho > 0.5$), the opinion disparity is greater when the community 1 adopts opinion 1 than when it adopts opinion 0, indicated by $\psi_1 > \psi_2$. In panel (2), we see that there are regimes for small enough $\epsilon_2$ where opinion disparity \textit{only} exists for imbalanced community structure.}
    \label{fig:rho_vs_epsilon2_psi}
\end{figure}

Changing the relative sizes of the communities not only changes the epidemic threshold, but it also affects the presence and strength of opinion disparity. To quantify opinion disparity in more detail, we define the one-sided opinion disparities to be
\begin{align}
\psi_1 &= \widetilde{x}_1 - \widetilde{x}_2\\
\psi_2 &= \widehat{x}_2 - \widehat{x}_1,
\end{align}
where $(\widetilde{x}_1, \widetilde{x}_2)$ is the stable fixed point of Eqs.~(\ref{eq:mf_imbalanced1})-(\ref{eq:mf_imbalanced2})
with the largest value of $x_1 - x_2$ and $(\widehat{x}_1, \widehat{x}_2)$ is the stable fixed point of Eqs.~(\ref{eq:mf_imbalanced1})-(\ref{eq:mf_imbalanced2})
with the largest value of $x_2 - x_1$. This allows us to more easily observe which fixed points disappear and appear as we vary $\rho$.

\begin{figure}
    \centering
    \includegraphics[width=8.6cm]{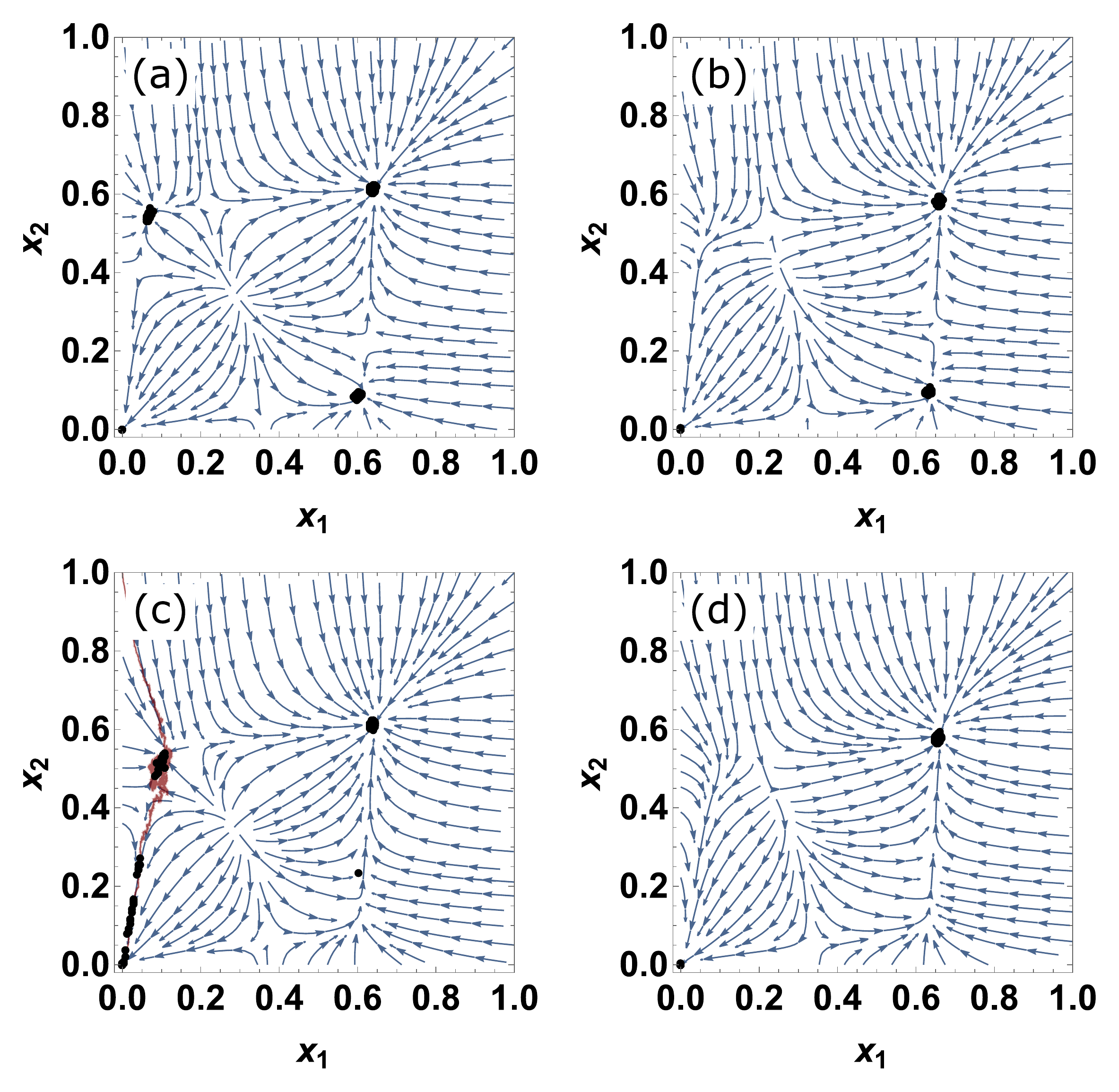}
    \caption{The phase plots corresponding to points (a)-(d) in Fig.~\ref{fig:rho_vs_epsilon2_psi} with the fixed points $(x_1, x_2)$ computed from numerical simulations overlaid on top. Each point is generated by computing the time-weighted average of the fraction of communities 1 and 2 infected for different initial conditions and hypergraph realizations, described more in Appendix~\ref{sec:appendix_simulations}. Five sample trajectories with the initial condition of $(x_1, x_2) = (0, 1)$ are shown in red in panel (c) to illustrate that fluctuations due to finite size effects can cause the state of the system to leave the asymmetric fixed point.}
    \label{fig:imbalanced_phase_plots}
\end{figure}

In Fig.~\ref{fig:rho_vs_epsilon2_psi} we illustrate how opinion disparity changes as the relative sizes of the communities are changed by slightly tuning the parameter $\rho$ away from $\rho = 1/2$. In the top panel we plot the quantity $\psi_1 - \psi_2$ as a function of $\rho$ and $\epsilon_2$. This quantity is positive (light colors) when the state with larger disparity has a larger value of $x_1$, and negative (dark colors) when it has a larger value of $x_2$. The amount and type of opinion disparity depend very sensitively on both the size imbalance parameter $\rho$ and the amount of pairwise community structure $\epsilon_2$. In panels (1) and (2) we show the one-sided opinion disparities $\psi_1$ and $\psi_2$ as a function of $\rho$ for two fixed values of $\epsilon_2$, corresponding to the two horizontal lines shown in the top panel. These figures show how asymmetric fixed points appear and disappear as the relative size of the communities changes, giving rise to the pattern in the top panel. The phase plots corresponding to points (a)-(d) marked with stars in the top panel of Fig.~\ref{fig:rho_vs_epsilon2_psi} are displayed in Fig.~\ref{fig:imbalanced_phase_plots}(a)-(d), overlaid with the numerically calculated fixed points and several reference trajectories. The fixed points are determined by first randomly infecting a fraction $\rho_1$ of communities 1 and $\rho_2$ of community 2 from a grid of initial conditions $(\rho_1, \rho_2)\in \{0,0.05,\dots,1\} \times \{0,0.05,\dots,1\}$ and simulating the system for $t=0,\dots,300$. Each black dot represents a weighted average of the last 10 time steps for each initial condition (more details are contained in Appendix~\ref{sec:appendix_simulations}). These numerical simulations show that the stable mean-field fixed points are generally a good representation of the stable states of the stochastic model. However, it is possible that, even though a fixed point is stable in the mean-field description, finite size effects can drive the stochastic model towards another fixed point. This is illustrated in Fig.~\ref{fig:imbalanced_phase_plots}(c), which shows five sample trajectories starting from the initial condition $(x_1,x_2) = (0,1)$. After hovering around the asymmetric fixed point, some of them drift towards the stable fixed point $(0,0)$. 

The phase plots in Fig.~\ref{fig:imbalanced_phase_plots} illustrate how the fixed points and their stability change as the imbalance parameter is varied. For small enough $\epsilon_2$ we see the appearance of regions where opinion disparity cannot occur for equal community sizes, but can occur for a small range of imbalanced community structures [e.g., see panel (c)]. In addition, for a large enough difference in community sizes, a majority of "1" opinions is untenable for the smaller community. The plot of $\psi_1 - \psi_2$ with respect to $\rho$ and $\epsilon_3$ is qualitatively similar so we omit it here. Increasing $\rho$ from the balanced case ($\rho = 0.5$) increases the size of the region where $\psi_2 > 0$ up to a critical value, after which it decreases. Likewise, the size of the region where $\psi_1 > 0$ decreases with increasing $\rho$. The same phenomenon occurs for $\epsilon_3$ with respect to $\rho$.

In Fig.~\ref{fig:imbalanced_phase_plots}, we see close agreement between the phase diagram corresponding to the Eqs.~\eqref{eq:mf_imbalanced1}-\eqref{eq:mf_imbalanced2} and the fixed points computed from numerical simulations.

Notably, regions with strong opposite opinion disparity types are adjacent: there are continuous trajectories in the $(\rho,\epsilon_2)$ phase space such that there is a discontinuous change of sign in the quantity $\psi_1-\psi_2$ along them. In practice this would mean that the establishment or destruction of a few inter-community ties, or a small change in community memberships, could result in a sudden change in opinion disparity.

\section{\label{sec:discussion} Discussion}

In this paper we studied the effect of community structure and higher-order interactions on the existence of opinion disparity in a simple collective social contagion model. To do so, we introduced a simple generative hypergraph model for hypergraphs with tunable community structure. We analyzed the dynamics of the contagion model using a mean-field approach and found that opinion disparity can be sustained for a relatively large range of community structure and infection rate parameters. In addition, we found that very small changes in the community sizes can lead to dramatic changes in the opinion disparity.

Here we discuss the limitations of our approach and possible directions for further research. First, our formalism is a model of discrete opinions, but in many situations opinions are better described as a continuous spectrum. In addition, opinions may not be one-dimensional, but rather multi-dimensional. As mentioned in Section~\ref{sec:opinion_model}, the contagion model we chose is asymmetric, which leads to behavior that one may not see if the two opinions were equivalent. Thus, our model is more appropriate to describe opinions or behavior that need to be sustained by group pressure, instead of describing a choice between two symmetric opinions. In addition, we assumed that the dynamics of the opinion formation process does not affect the underlying structure of the hypergraph. The formation of echo chambers, however, may be the result of not only the dynamics of opinion formation but also of the hypergraph adapting to dissolve interactions between individuals with dissimilar opinions \cite{kan_adaptive_2023}. For simplicity, we considered hypergraphs with only two communities and interactions of sizes 2 and 3. However, our models and methods could be extended to model any number of communities and hyperedge sizes. In addition, we only considered variants of the stochastic block model with homogeneous degree, but empirical higher-order systems often contain degree heterogeneity. Including this heterogeneity in our hypergraph models could help bridge the gap between theory and empirical datasets. To this end, it would be useful to simulate the contagion model that we have described on empirical data sets to see if opinion disparity occurs, similar to the approach of Ref.~\cite{ferraz_de_arruda_multistability_2023}. Lastly, another drawback of our approach is that we need the community labels of the nodes to compute the opinion disparity, but ground truth labels are not always available. To alleviate this, one could complement our approach with community detection algorithms.

Despite these limitations, our work shows that the community structure of higher-order interactions may be one of the ingredients contributing to opinion disparity. In contrast to models of opinion formation which require opinion homophily for opinion disparity to exist, our model only requires the presence of higher-order interactions and sufficiently disconnected communities for opinion disparity to occur. Thus, our results should be interpreted as the exploration of a complementary mechanism of opinion disparity formation. Our results also show that communities of differing sizes can modify the epidemic threshold and create regimes where opinion disparity can occur despite being impossible for equally-sized communities.

\section*{Data Availability Statement} The code and datasets supporting this work are openly available on \href{https://github.com/nwlandry/opinion-disparity-in-hypergraphs}{GitHub} \cite{landry_code_2023}.

\acknowledgments{Nicholas Landry acknowledges financial support from the National Science Foundation Grant 2121905, "HNDS-I: Using Hypergraphs to Study Spreading Processes in Complex Social Networks", and from the National Institutes of Health 1P20 GM125498-01 Centers of Biomedical
Research Excellence Award. JGR acknowledges support from NSF Grant DMS-2205967. Nicholas Landry would like to acknowledge helpful feedback from Mari Kawakatsu on the conception of this project. }

\appendix

\section{\label{sec:appendix_hsbm_algorithm} An efficient algorithm for sampling m-HSBM hypergraphs}

Simply sampling na\"ively from the list of all possible hyperedges of size $m$ and accepting them with probability $p \ll 1$ has complexity \order{N^m} which can be computationally prohibitive for large $m$. We modify the algorithm presented in Ref.~\cite{miller_efficient_2011} to $m$-uniform hypergraphs to sample from the m-HSBM with time complexity \order{m(N+|E|)}.

In Section \ref{sec:ppm} it is implicitly assumed that no multiedges or self-loops may occur. However, given an index in the list of all possible unique hyperedges, it can be expensive to recover the hyperedge corresponding to that index. If instead we allow these artifacts to occur, then it is \order{m} to recover the hyperedge of interest in contrast to iterating through all unique combinations, which is \order{mN}. Because of this modification, in practice, we divide every probability derived in Section~\ref{sec:ppm} by $m!$ to account for the increase in possible hyperedges because of multiedges and hyperedges containing self-loops. Because allowing multiple instances of the same node changes the number of unique neighbors, we remove these artifacts when they occur.

Consider an $m$-uniform hypergraph with $N$ nodes and a community label $g_i$ for each node $i$. We denote by ${\bf g}$ the vector of all node labels, i.e., ${\bf g}=[g_1,\dots,g_N]$. The number of unique community labels is $G$ and, as in Section~\ref{sec:hsbm}, $P_{g_{i_1}, \dots, g_{i_m}}$ specifies the probability that nodes $i_1, \dots, i_m$ with community labels $g_{i_1}, \dots, g_{i_m}$ form a hyperedge. The function $\theta(g)$ returns a vector of all the nodes that have community label $g$ and $|\theta(g)|$ is the number of nodes with community label $g$. We iterate through each entry $b_1, \dots, b_m$ of the tensor $P$ and for fixed group assignments for the nodes in a hyperedge, the probability of generating a hyperedge is constant. We generate all the hyperedges associated with these ordered community assignments. The hyperedges are elements of the set formed by the Cartesian product of the indices in each partition and the maximum index is given by the product $\prod_{b\in {\bf b}} |\theta(b)|$.

\begin{algorithm}
    \caption{Return an $m$-hyperedge from a specified index, given community partitions (IndexToEdge)}
    \label{alg:getedge}
    \DontPrintSemicolon
    \KwIn{$i$, ${\bf b}$, $\theta$}
    \KwOut{$e$}
    $e=\emptyset$\;
    $m = |{\bf b}|$\;
    $r = m$\;
    \While{$r > 0$}
    {
        $j = \left\lfloor i/\left(\prod_{p=r+1}^{m} |\theta(b_r)|\right)\right\rfloor \mod |\theta(b_r)|$\;
        $v = \theta(b_r)_j$\;
        $e \leftarrow e \cup v$\;
        $r \leftarrow r - 1$\;
    }
    \KwRet $e$\;
\end{algorithm}

Our algorithm is an extension of the algorithm in Ref.~\cite{miller_efficient_2011} and the main idea is this: instead of iterating through all possible edges and accepting an edge with probability $P_{g_{i_1}, \dots, g_{i_m}} = p$, which is expensive when $p \ll 1$, we simply skip the edges that would be rejected by sampling from a geometric distribution. While the current index is less than the maximum index, we increment the index with steps $s \sim \text{Geometric}_1(p)$, the distribution of the number of Bernoulli trials needed for a success. For a given index, we convert to a list of $m$ node labels with Algorithm~\ref{alg:getedge}. Because we simulate the community connection probability tensor patch-by-patch, for a given patch, we specify the community to which each node belongs as an ordered list. The algorithm for sampling from the m-HSBM is given in Algorithm~\ref{alg:sbm}.

\begin{algorithm}
    \caption{Generating the $m$-uniform stochastic block model for hypergraphs (m-HSBM)}
    \label{alg:sbm}
    \DontPrintSemicolon
    \KwIn{${\bf g}$, $m$, $P$}
    \KwOut{$E$}
    $N = |{\bf g}|$\;
    $G = |\text{unique }b \in {\bf g}|$\;
    $B = 1, \dots, G$\;
    $\theta : b \mapsto (i \ | \ g_i = b, \ i = 1,\dots, N)$\;
    $E = \emptyset$\;
    \For{${\bf b}=(b_1,\dots,b_m) \in B \times \dots \times B$}
    {
        $M=\prod_{b\in {\bf b}} |\theta(b)|$\;
        $p = P_{b_1,\dots,b_m}$\;
        $i \sim \text{Geometric}_1(p)$\;
        \While{$i < M$}
        {
            $e = \text{IndexToEdge}(i, {\bf b}, \theta)$\;
            \If{$|e| = m$}
            {
                $E \leftarrow E \cup e$\;
            }
            $i \leftarrow i + \text{Geometric}_1(p)$\;
        }
    }
    \KwRet $E$\;
\end{algorithm}

Generating an \ER hypergraph is a special case of Algorithm~\ref{alg:sbm} where there is a single community ($G=1$, $g_i=1, \ i=1\dots N$, $|\theta({\bf g})|=N$) and an efficient algorithm is provided in Ref.~\cite{landry_code_2023}.

\section{\label{sec:appendix_simulations} Numerical simulations} We use the same Gillespie algorithm described in Ref.~\cite{landry_hypergraph_2022} to simulate the hypergraph SIS model efficiently. We obtain the community labels of the nodes from the labels specified in our m-HPPM model. We use these labels to return the number of nodes infected in each community with respect to time based on the community labels of the nodes. There are three stable states that the simulation can reach in equilibrium: first, the fractions of communities 1 and 2 can remain well-separated indicating the possibility of opinion disparity; second, the epidemic equilibrium where the average fraction infected is identical in communities 1 and 2; and third, where the simulation dies out and there are no infected individuals in either community 1 or community 2. We assume that if we start in the $({\bf x}_1={\bf 1}, {\bf x}_2={\bf 0})$ state and the structural and dynamical parameters admit a polarized stable state, then $x_1$ and $x_2$ will remain well-separated. We specify that the initial state of every node in community 1 is infected and every node in community 2 is susceptible and run the simulation until a maximum time of $t_{max} = 100$ is reached or every node is in the susceptible state. There is a non-zero probability, however, that the second or third case will occur for a weakly stable asymmetric fixed point. We heuristically tuned the simulation time to minimize the effect of finite size effects. We take the absolute value of the difference between these two time series to obtain the opinion disparity as a function of time and perform time-weighted averaging of the resulting time-series from the last 10\% of the time series (described in more detail in Ref.~\cite{landry_hypergraph_2022}). For each value of $\epsilon_2$ and $\epsilon_3$, we generate a single realization from the m-HPPM and for each set of infectious parameter values, we run a single simulation to preserve the sharp transitions between regions that admit opinion disparity and those that do not.

When generating the fixed points overlaying the phase diagrams in Fig.~\ref{fig:imbalanced_phase_plots}, we simulate the contagion dynamics for many different initial states. We generate the initial nodal states by iterating over a grid of initial infection densities $(\rho_1, \rho_2) \in \{0, 0.05, \dots, 1\} \times \{0, 0.05, \dots, 1\}$. For each $(\rho_1, \rho_2)$ combination, we sample uniformly at random $\rho_1 \rho N$ nodes from community 1 and $\rho_1 (1- \rho)N$ nodes from community 2. We set the state of these nodes to be infected and simulate the contagion process with this initial condition up to $t_{max}=300$. We generate a single m-HPPM hypergraph with $\langle k\rangle = \langle q\rangle = 50$, and for each initial condition, we simulate the contagion process. We perform time-weighted averaging on the states of the nodes in communities 1 and 2 as described above.

\section{\label{sec:appendix_stability} Calculating the stability of fixed points}

Here, we present the Jacobian, $J(x_1, x_2)$, for the hypergraph SIS model on both the planted partition model and the imbalanced planted partition model.

The Jacobian for the system of equations Eqs.~\eqref{eq:mf1_rescaled}-\eqref{eq:mf2_rescaled} governing contagion spread on the planted partition model is given by
\begin{align}
J_{1,1} = & -1 + \frac{\widetilde{\beta}_2}{2}(1 - x_1)\left[1 + \epsilon_2\right]\nonumber\\
& - \frac{\widetilde{\beta}_2}{2}[x_1 + x_2 + \epsilon_2(x_1 - x_2)]\nonumber\\
& + \frac{\widetilde{\beta}_3}{2}(1 - x_1)[x_1 + x_2 + \epsilon_3 (3 x_1 - x_2)]\nonumber\\
& - \frac{\widetilde{\beta}_3}{4}[(x_1 + x_2)^2 + \epsilon_3(3 x_1^2 - 2 x_1 x_2 - x_2^2)],
\end{align}

\begin{align}
J_{1,2} = & \frac{\widetilde{\beta}_2}{2} (1 - x_1)[1 - \epsilon_2]\nonumber\\
& + \frac{\widetilde{\beta}_3}{2}(1 - x_1)[x_1 + x_2 - \epsilon_3 (x_1 + x_2)].
\end{align}
$J_{2,2}$ and $J_{2,1}$ can be obtained by noting the symmetry of Eqs.~\eqref{eq:mf1_rescaled}-\eqref{eq:mf2_rescaled} and substituting $x_1 \leftrightarrow x_2$ into the expressions for $J_{1,1}$ and $J_{1,2}$.

For the system of equations Eqs.~\eqref{eq:mf_imbalanced1}-\eqref{eq:mf_imbalanced2} governing the contagion dynamics of the imbalanced planted partition model, the Jacobian is 
\begin{align}
 J_{1,1} &= -1 + \beta_2 \langle k\rangle \rho (1 - x_1)(1 + r_{\rho,2}\, \epsilon_2)\nonumber\\
& - \beta_2 \langle k\rangle [(1 - \rho)(1 - \epsilon_2) x_2 + \rho (1 + r_{\rho,2}\, \epsilon_2) x_1]\nonumber\\
& + 2 \beta_3\langle q\rangle(1 - x_1)[\rho^2 (1 + r_{\rho,3}\,\epsilon_3) x_1\nonumber\\
& + (1 - \rho) \rho (1 - \epsilon_3)x_2]\nonumber\\
& - \beta_3\langle q\rangle (\rho^2 (1 + r_{\rho,3} \epsilon_3)x_1^2\nonumber\\
& + 2\rho(1 - \rho)(1 - \epsilon_3) x_1 x_2 + (1 - \rho)^2 (1 - \epsilon_3) x_2^2)
\end{align}

\begin{align}
 J_{1,2} &= \beta_2\langle k\rangle (1 - \rho)(1 - \epsilon_2)(1 - x_1)\nonumber\\
& + 2\beta_3\langle q\rangle (1 - \rho)\rho (1 - \epsilon_3)(1 - x_1)x_1\nonumber\\
& + 2\beta_3\langle q\rangle(1 - \rho)^2 (1 - \epsilon_3)(1 - x_1) x_2
\end{align}
$J_{2,1}$ and $J_{2,2}$ can be calculated by noting that Eq.~\eqref{eq:mf_imbalanced2} may be obtained from Eq.~\eqref{eq:mf_imbalanced1} by substituting $x_1 \leftrightarrow x_2$ and $\rho \leftrightarrow (1 - \rho)$.

\bibliography{references,code_references}

\end{document}